\documentstyle[12pt,epsfig]{article}

\newcommand{\gs}{\gamma^*}

\newcommand{\sigmahat}{\hat{\sigma}}
\newcommand{\shat}{\hat{s}}
\newcommand{\that}{\hat{t}}
\newcommand{\uhat}{\hat{u}}
\newcommand{\eps}{\epsilon}

\newcommand{\pup}{p^{\uparrow}}

\newcommand{\Sp}{S_{\perp}}
\newcommand{\la}{\langle}
\newcommand{\ra}{\rangle}
\newcommand{\GFt}{\widetilde{G}_F}
\newcommand{\GFa}{G_{F,a}}
\newcommand{\GFta}{\widetilde{G}_{F,a}}

\newcommand{\bea}{\begin{eqnarray}}
\newcommand{\eea}{\end{eqnarray}}
\newcommand{\nn}{\nonumber}

\newcommand{\bfc}{\begin{figure}\begin{center}}
\newcommand{\efc}{\end{center}\end{figure}}

\newcommand{\fig}[2]{\scalebox{#1}{\includegraphics{#2}}}

\newcommand{\Slash}[1]{\ooalign{\hfil/\hfil\crcr$#1$}}

\begin{document}

\vspace*{0.5cm}

\begin{center}

{\Large \bf Hard- and soft-fermion-pole contributions to single \\[3mm]
transverse-spin asymmetry for Drell-Yan process}

\vspace{1cm}

Koichi Kanazawa$^1$ and Yuji Koike$^2$

\vspace{1cm}

{\it $^1$ Graduate School of Science and Technology, Niigata University,
Ikarashi, \\ Niigata 950-2181, Japan}

\vspace{0.5cm}

{\it $^2$ Department of Physics, Niigata University,
Ikarashi, Niigata 950-2181, Japan}

\vspace{2.5cm}

{\large \bf Abstract} \end{center}

We study the single transverse-spin asymmetry for the Drell-Yan 
lepton pair production 
based on the twist-3 mechanism in the collinear factorization. 
We calculate all the hard-pole (HP) and the soft-fermion-pole (SFP)
contributions to the single-spin-dependent cross section
originating from the quark-gluon correlation functions
in the transversely polarized nucleon
in the leading order with respect to the QCD coupling constant.  
Combined with the soft-gluon-pole (SGP) contribution, 
this completes the corresponding twist-3 cross section.  
In the real photon limit, where all the HP contributions are transformed 
into the SFP contribution, we find that 
the SFP partonic hard cross section for the two independent quark-gluon correlation functions
coincides in each scattering channel, as in the case of the inclusive
light-hadron production.  Our result enables one to extract the
quark-gluon correlation functions 
from the forthcoming experiments 
at several facilities such as 
RHIC and J-PARC.

\newpage

Large single transverse-spin asymmetries (SSA) observed in hard inclusive 
processes have been recognized as a key milestone to probe
the internal structure of hadrons beyond the collinear
parton model (see \cite{review} for a recent review).  
On the experimental side, SSA's in inclusive hadron
production in pp collisions\,\cite{E7041991}-\cite{Phenix2010}
and semi-inclusive deep-inelastic scattering (SIDIS)\,\cite{Compass2005}-\cite{Hermes2009}
have been widely studied in the last decades, and a wealth of SSA data has
been reported for the production of a variety of hadrons 
in a wide kinematical domain. 
Most of the SSA data in SIDIS is in the region of the small transverse momentum 
of the final hadron and thus has been analyzed
within the framework of the transverse-momentum-dependent 
factorization\,\cite{CS81,CSS85,JMY05}.   
In this framework, correlations between the intrinsic transverse momentum of partons and the spin vector
causes SSA and they are classified into the 
Sivers\,\cite{Sivers1990,Sivers1991,Collins2002} and
Collins\,\cite{Collins93} effects, which are
associated with the initial- and final-state hadrons,
respectively. 
In the SIDIS kinematics, these two effects appear with different dependences on the azimuthal angles
and there have been some attempts to extract responsible functions from the analysis
of the existing data\,\cite{AnselminoEtal2007,AnselminoEtal2009}. 
On the other hand, 
for the inclusive single hadron production
in pp collision with large transverse momentum, one can analyze the process in the
framework of the collinear factorization, 
in which SSA is described as a leading twist-3 observable
originating from multi-parton correlations\,\cite{QiuSterman1992}-\cite{KangYuanZhou2010}.  
In the twist-3 mechanism, 
multi-parton correlations are also represented by the twist-3 multi-parton
correlation functions in the initial nucleon\,\cite{QiuSterman1992}-\cite{Koike:2011mb} 
and the twist-3 fragmentation functions
for the final hadron\,\cite{YZ09,KangYuanZhou2010}.  For the SSA in the single hadron production
in pp collisions,
these two effects appear with the same angular dependence and can not be separated kinematically.  
Although RHIC $A_N$ data for $\pup p\to hK$ ($h=\pi,K$)
has been well described by assuming solely the effect of the 
quark-gluon correlation functions
in the polarized nucleon\,\cite{KouvarisQiuVogelsangYuan2006,KanazawaKoike2010,KanazawaKoike2011}, 
relevance of the description is yet to be clarified by a global analysis including
the effects of the multi-gluon correlation functions\,\cite{Beppu:2010qn,Koike:2011ns,Koike:2011mb} 
and the twist-3 fragmentation function\,\cite{YZ09,KangYuanZhou2010}.

The SSA in the
Drell-Yan (DY) lepton pair production, $\pup p\to \gamma^* X\to \ell^+\ell^- X$, 
and the direct-photon 
production, $\pup p\to \gamma X$, 
provides us with a unique opportunity to determine the twist-3 quark-gluon
correlation functions in the nucleon, since there is no fragmentation ambiguity
in the final state.  
Owing to the naively $T$-odd nature of SSA, it occurs from an interference 
between the amplitudes which have different complex phases.  
In the twist-3 mechanism, this phase is supplied as a pole contribution from an internal
propagator in the hard part.  These poles fix the momentum fraction
carried by one of the parton lines emanating from the
parent nucleon.   Corresponding to the cases in which (i) gluon line becomes soft, 
(ii) one of the fermion lines becomes soft, and (iii) none of the
parton lines becomes soft, 
those poles are classified, respectively, as the soft-gluon-pole
(SGP), the soft-fermion-pole (SFP) and the hard-pole (HP).  
For the SSA in the DY process, all these poles contribute to the
single-spin-dependent cross section.  
Among them, the contributions from SGP and a part of HP 
have been calculated in \cite{JiQiuVogelsangYuan2006DY,KT071}, while  
there has been no calculation for the SFP contribution.
Furthermore, some diagrams for the HP contribution
were overlooked in the previous calculation.  
Existence of the new type of diagrams in which a quark-antiquark pair resides in the same side of the 
final-state cut
was recognized earlier in the study of SSA in SIDIS
for the SFP
contribution\,\cite{KVY08} and for the HP contributions\,\cite{Koike:2009yb}, 
and also in the study on the SFP
contribution to $\pup p\to \pi X$\,\cite{KoikeTomita2009}.  
For the direct photon production, in which HP's merge into SFP's,
the contribution from those diagrams was not considered in the previous study\,\cite{QiuSterman1992}.  
Given the unique role of SSA in the DY and the direct-photon processes
as stated above,
it is necessary and important to have the complete twist-3 section formula for these processes.

The purpose of this Letter is to derive the entire HP and SFP
contributions associated with the twist-3 quark-gluon correlation functions
to these processes.  
Combining the result with the known SGP cross section,
this will complete the twist-3 cross sections and 
will enable us to extract more complete and pure information on the
quark-gluon correlations from experiments.

We consider the SSA for the DY process 
\bea
\pup (p,\Sp) + p (p') \to \gs (q) + X, 
\eea
where $\gs$ is the virtual photon with the four momentum $q$ ($q^2=Q^2$) decaying into a lepton pair 
in the final
state. $p$ and $p'$ are the four momenta of the initial protons which can be regarded as lightlike
($p^2=p'^2=0$) in the twist-3 accuracy, 
and $\Sp$ is
the spin vector of the transversely polarized proton 
normalized as $\Sp^2 = -1$. 
In the transversely
polarized proton there are two independent twist-3 quark-gluon correlation functions, 
$G_{F,a}(x_1,x_2)$ and $\GFta(x_1,x_2)$, for each quark flavor $a$ 
defined as
\bea
M_{Fij}^{\alpha} (x_1,x_2) &=& \int \frac{d\lambda}{2\pi} \int
\frac{d\mu}{2\pi} e^{i\lambda x_1}
e^{i\mu(x_2-x_1)} \la p\Sp | \bar{\psi}_j^a(0) g F^{\alpha\beta}(\mu n)
n_{\beta} \psi_i^a(\lambda n) | p\Sp \ra \nn\\
&=& \frac{M_N}{4}(\Slash{p})_{ij} \eps^{\alpha pn\Sp}
G_{F,a}(x_1,x_2) +
i\frac{M_N}{4} (\gamma_5\Slash{p})_{ij} \Sp^{\alpha}
\GFta (x_1,x_2) + \cdots ,
\label{MF}
\eea
where $\psi_i^a$ is the quark field with spinor index $i$,
$F^{\alpha\beta}$ is the gluon's field strength and the gauge link
operators between these fields are suppressed for simplicity.  
$n$ is another lightlike vector satisfying $p\cdot n=1$ and $n\cdot
\Sp=0$, and 
$\eps^{\alpha pn\Sp} \equiv \eps^{\alpha\mu\nu\rho} p_{\mu} n_{\nu}
S_{\perp\rho}$ with $\eps_{0123} = 1$.  
$M_N$ is the nucleon mass introduced to define the correlation functions dimensionless
and the ellipses represent twist-4 or higher.  
The QCD coupling constant $g$ follows our convention for the covariant derivative
$D^{\mu} = \partial^{\mu}-igA^{\mu}$.  The variables $x_{1,2}$
and $x_2-x_1$ denote the longitudinal momentum fractions of the quarks
and the gluon coming out of the transversely polarized proton.  
%
%
We remind the symmetry property $G_{F,a}(x_1,x_2)=G_{F,a}(x_2,x_1)$ and 
$\widetilde{G}_{F,a}(x_1,x_2)=-\widetilde{G}_{F,a}(x_2,x_1)$.
(See \cite{EguchiKoikeTanaka2006,EguchiKoikeTanaka2007} for the detailed property of
$G_F(x_1,x_2)$ and $\GFt(x_1,x_2)$.) 
Following our convention in \cite{KoikeTomita2009}, 
the correlation functions for the anti-quark flavor 
are related to those for the 
quark flavor as
\begin{eqnarray}
G_{F,\bar{a}}(x_1,x_2)=G_{F,a}(-x_2,-x_1),\qquad 
\widetilde{G}_{F,\bar{a}}(x_1,x_2)=-\widetilde{G}_{F,a}(-x_2,-x_1).
\end{eqnarray}
We also remind that the replacement of the field strength by the covariant derivative as
$g F^{\alpha\beta}(\mu n)n_{\beta}\to D^\alpha (\mu n)$ in (\ref{MF})
defines other twist-3 quark-gluon correlation functions $G_{D,a}(x_1,x_2)$ and $\widetilde{G}_{D,a}(x_1,x_2)$
in the same decomposition corresponding to $G_{F,a}(x_1,x_2)$ and $\widetilde{G}_{F,a}(x_1,x_2)$,
respectively.  However, they are not independent, in particular, at $x_1\neq x_2$, which are relevant to
the HP and SFP contributions, they are related by the relation\,\cite{EguchiKoikeTanaka2006}, 
\bea
G_{D,a}(x_1,x_2)={G_{F,a}(x_1,x_2)\over x_1-x_2},\qquad   
\widetilde{G}_{D,a}(x_1,x_2)={\widetilde{G}_{F,a}(x_1,x_2)\over x_1-x_2},\qquad (x_1\neq x_2).
\eea

\begin{figure}[t]
 \begin{center}
  \fig{0.5}{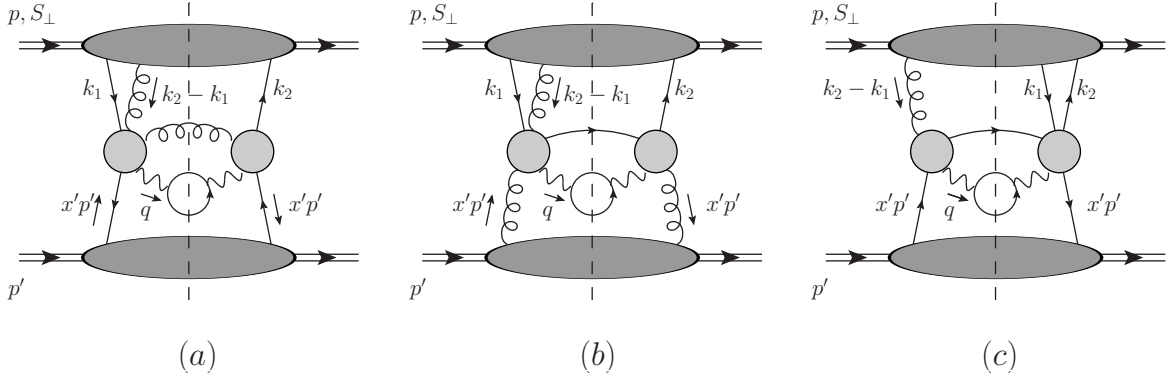}
 \end{center}
 \caption{Generic diagrams giving rise to the twist-3 
single-spin-dependent cross section for the Drell-Yan process originating from the quark-gluon correlation 
functions in the nucleon (upper blob).  Momenta of the parton lines coming out of the
polarized nucleon are set to $k_i=x_ip$ ($i=1,2$).  
Mirror diagrams also contribute. \label{general}}
\end{figure}

Figure~\ref{general} shows the generic twist-3 diagrams giving rise to 
SSA for the DY process.  Their mirror diagrams also give the same contribution
as those shown in Fig. \ref{general}.  
Following the twist-3 formalism for the contribution of the quark-gluon correlation
functions developed in \cite{EguchiKoikeTanaka2007} (see also \cite{KoikeTomita2009}), 
one can obtain 
the single-spin-dependent cross section from the HP and SFP as
\bea
\frac{d^4\Delta\sigma^{\rm HP, SFP}}{dQ^2dyd^2\vec{q}_{\perp}} &=&
\frac{\alpha_{em}^2 \alpha_s}{3\pi S Q^2}
 \int\frac{dx'}{x'}
\int dx_1 \int dx_2 \frac{1}{x_1-x_2} \nn\\
&& \times {\rm Tr}[i
M_F^{\alpha} (x_1,x_2) S_{\alpha}^{\rm HP, SFP} (x_1p,x_2p,x'p',q)] f(x'),
\eea
where $y$ and $\vec{q}_{\perp}$ are, respectively, the rapidity and the transverse momentum of
the virtual photon, $S=(p+p')^2$ is the center of mass energy,
$\alpha_{em}\simeq 1/137$ and
$\alpha_s=g^2/(4\pi)$ are, respectively, the electromagnetic and the strong coupling constants,
$f(x')$ is the unpolarized parton density in the
unpolarized proton.  
$S_{\alpha}^{\rm HP, SFP} (x_1p,x_2p,x'p',q)$ is the corresponding hard part 
for the HP and SFP contributions in which parton momenta coming from the polarized nucleon
are set to $k_i=x_ip$ ($i=1,2$).  
${\rm Tr}[\cdots ]$ indicates the trace over spinor and color indices.

\begin{figure}[t]
 \begin{center}
  \fig{0.35}{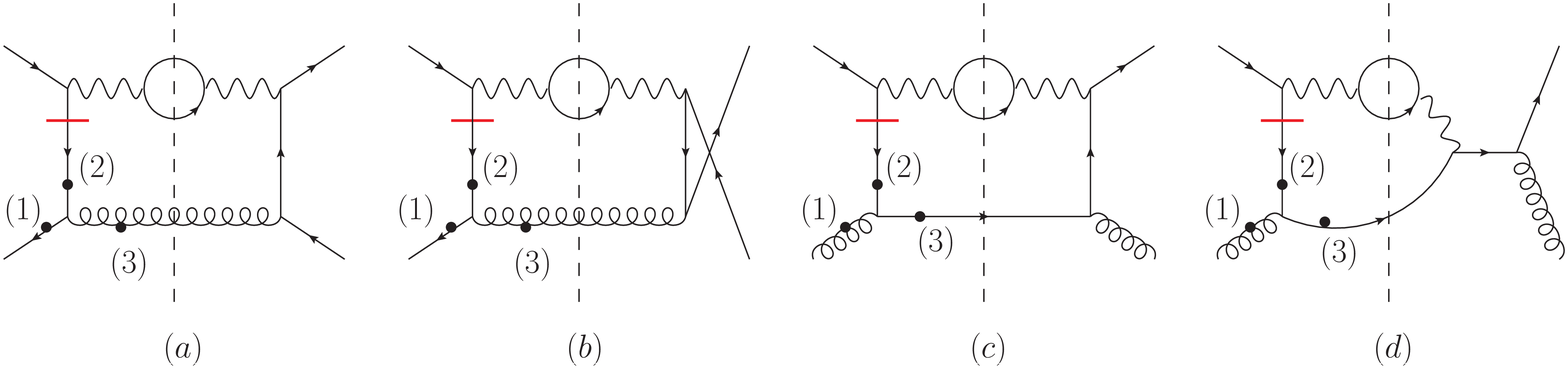}
 \end{center}
\caption{Feynman diagrams for the HP contribution (HP1) 
corresponding to Figs. 1 (a) and (b).  An extra coherent gluon line coming out of the polarized nucleon
attaches to one of the dots numbered as (1), (2) and (3).  The propagator with a short bar gives
rise to the HP at $x_1 = x_B$. 
\label{HP1}}

 \begin{center}
  \fig{0.35}{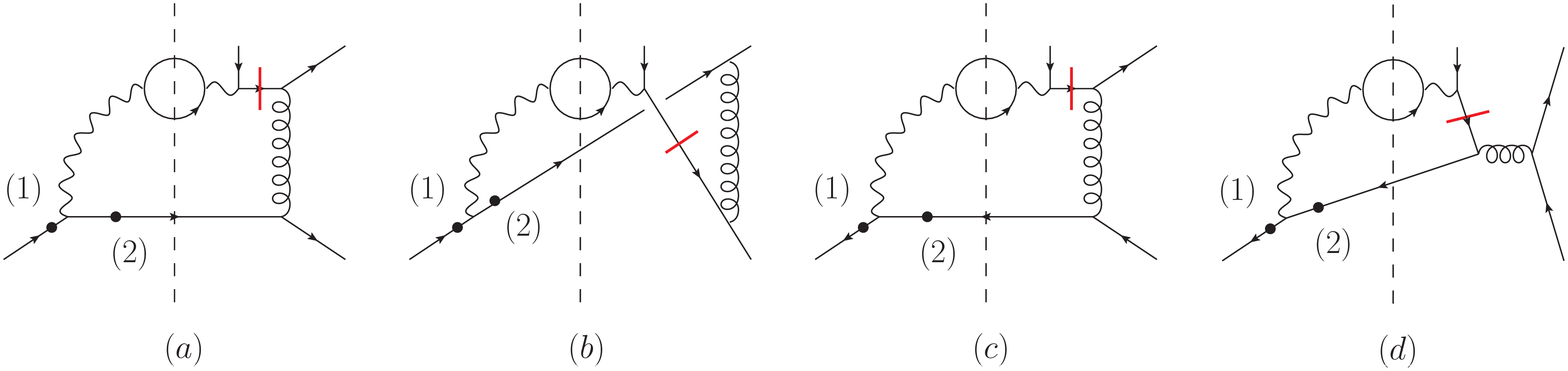}
 \end{center}
 \caption{Feynman diagrams for the HP contribution (HP2) 
corresponding to Fig. 1 (c).  An extra coherent gluon line coming out of the polarized nucleon
attaches to one of the dots numbered as (1) and (2).  The propagator with a short bar gives
rise to the HP at $x_1 = -x_B$. 
\label{HP2}}
\end{figure}

Figures~\ref{HP1} and \ref{HP2} show the two types of diagrams generating HP's.
Those in Fig. \ref{HP1} were considered for the $G_F$ contribution in \cite{JiQiuVogelsangYuan2006DY}.  
In these diagrams, the quark propagator with a short bar carrying the momentum
$x_1p-q$ gives rise to the HP at $x_1 = Q^2/(2p\cdot q)\equiv x_B$.  
Diagrams in Fig. \ref{HP2} are the new ones not considered in \cite{JiQiuVogelsangYuan2006DY}
and have the two quark lines from the polarized nucleon in the same side of the
final-state cut.  There
the quark propagator with a short bar carrying the momentum
$x_1p+q$ gives rise to the HP at $x_1 =- x_B$.
Calculating the diagrams in Figs.~\ref{HP1} and \ref{HP2}, 
we obtain the HP contribution to the twist-3 cross section as
\bea
\frac{d^4\Delta\sigma^{\rm DY,HP}}{dQ^2dyd^2\vec{q}_{\perp}} &=&
\frac{\alpha_{em}^2 \alpha_s}{3\pi N S Q^2}
\frac{(-\pi) M_N}{2}\eps^{qpn\Sp} \int\frac{dx'}{x'} \int \frac{dx}{x}
\delta (\shat+\that+\uhat-Q^2) \nn\\
& \times & \sum_a \left[  
e_a^2 \left\{ \sigmahat^{\rm HP1}_{V, a\bar{a}} G_{F,a}(x_B,x) +
\sigmahat^{\rm HP1}_{A, a\bar{a}} \GFta(x_B,x) \right\} f_{\bar{a}}(x') \right.\nn\\
&& \quad + 
e_a^2 \left\{ \sigmahat^{\rm HP1}_{V, ag} G_{F,a}(x_B,x) +
\sigmahat^{\rm HP1}_{A, ag} \GFta(x_B,x) \right\} G(x') \nn\\
&& + \sum_{b} e_a e_b \left\{ \sigmahat^{\rm HP2}_{V, ab} G_{F,a}(-x_B,x-x_B)
+ \sigmahat^{\rm HP2}_{A, ab} \GFta(-x_B,x-x_B) \right\} f_b(x') \nn\\
&& + \left. \sum_{b} e_a e_b \left\{ \sigmahat^{\rm HP2}_{V, a\bar{b}}
 G_{F,a}(-x_B,x-x_B) 
+ \sigmahat^{\rm HP2}_{A, a\bar{b}} \GFta (-x_B,x-x_B) \right\} f_{\bar{b}}(x') 
\right], 
\label{HPcross}
\eea
where $e_a$ stands for the fraction of the electric charge for the quark flavor $a$,
$f_b(x')$ is the unpolarized quark density, and $G(x')$ is the unpolarized gluon density.  
In (\ref{HPcross}), the sum for $a$ is over all
quark and anti-quark flavors $(a=u,d,s,\bar{u},\bar{d},\bar{s}, ...)$, and 
$\sum_{b}$ indicates that the sum for $b$ is restricted over the quark flavors when $a$ is a quark
and over anti-quark flavors when $a$ is an anti-quark.  
The partonic hard cross sections are classified by the upper indices HP1 and HP2
corresponding, respectively, to Fig. \ref{HP1} 
and Fig. \ref{HP2}.  
They
are the functions of the 
Mandelstam variables
in the parton level, 
$\shat=(x p + x' p')^2$, $\that=(xp-q)^2$
and $\uhat=(x'p'-q)^2$ and the Casimir operators 
of SU(3) $C_F=(N^2-1)/(2N)$ and $T_R=1/2$ with the number of
colors $N=3$, and are given by 
\bea
&& \sigmahat^{\rm HP1}_{V, a\bar{a}} = 
\frac{ 4[(Q^2-\that)^3+Q^2\shat^2] }{ \that^2\uhat^2 }
\left[\frac{1}{2N}+C_F\frac{\shat}{Q^2-\that}\right] ,\label{JQVY1}\\
&& \sigmahat^{\rm HP1}_{A, a\bar{a}} = 
\frac{ 4[(Q^2-\that)^3-Q^2\shat^2] }{ \that^2\uhat^2 }
\left[\frac{1}{2N}+C_F\frac{\shat}{Q^2-\that}\right] ,\\
&& \sigmahat^{\rm HP1}_{V, ag} =
\frac{ 4[(Q^2-\that)^3+Q^2\uhat^2] }{ -\shat\that^2\uhat }
\left[\frac{-N^2}{2(N^2-1)}+T_R\frac{\shat}{Q^2-\that} \right] , \label{JQVY2}\\
&& \sigmahat^{\rm HP1}_{A, ag} =
\frac{ 4[(Q^2-\that)^3-Q^2\uhat^2] }{ -\shat\that^2\uhat}
\left[\frac{-N^2}{2(N^2-1)}+T_R\frac{\shat}{Q^2-\that} \right] , \\
&& \sigmahat^{\rm HP2}_{V, ab} =
\frac{2(Q^2+\that)}{\that\uhat(Q^2-\that)} 
\left[ \frac{\shat^2+\uhat^2}{Q^2-\that}
+ \frac{\shat^2}{N\uhat} \delta_{ab} \right],\\
&& \sigmahat^{\rm HP2}_{A, ab} =
\frac{-2}{\that\uhat}
\left[ \frac{\shat^2+\uhat^2}{Q^2-\that}
+ \frac{\shat^2}{N\uhat} \delta_{ab} \right],\\
&& \sigmahat^{\rm HP2}_{V, a\bar{b}} = 
\frac{-2(Q^2+\that)}{\that\uhat(Q^2-\that)} 
\left[ \frac{\shat^2+\uhat^2}{Q^2-\that}
+ \frac{\uhat^2}{N\shat} \delta_{ab} \right],\\
&& \sigmahat^{\rm HP2}_{A, a\bar{b}} = 
\frac{2}{\that\uhat}
\left[ \frac{\shat^2+\uhat^2}{Q^2-\that}
+ \frac{\uhat^2}{N\shat} \delta_{ab} \right]. 
\eea
The above results (\ref{JQVY1}) and (\ref{JQVY2}) agree with those in \cite{JiQiuVogelsangYuan2006DY},
while all the others are new.  

\begin{figure}
 \begin{center}
  \fig{0.35}{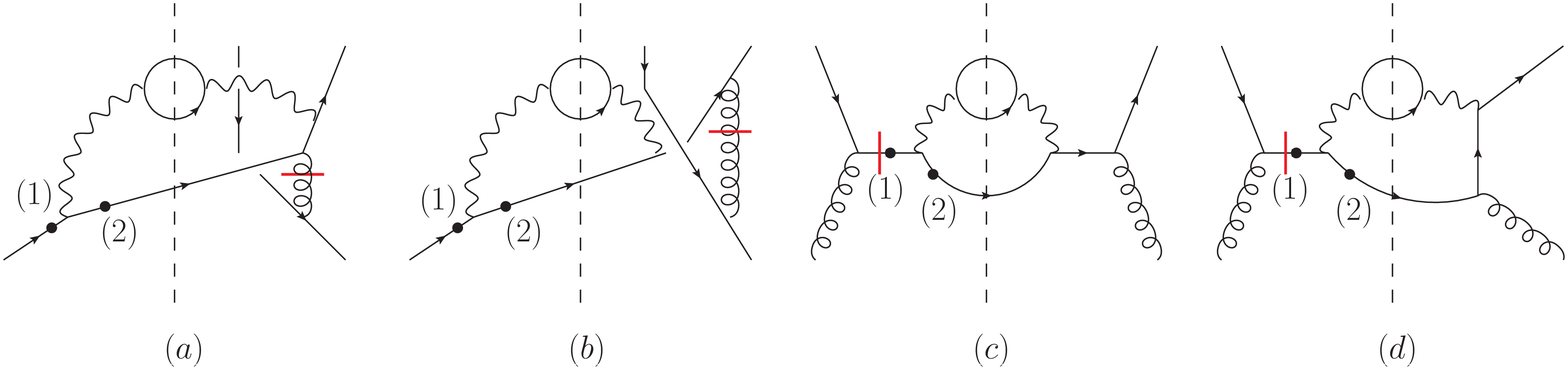}
 \end{center}
 \caption{
Feynman diagrams for the SFP contribution
corresponding to Figs. 1 (c) and (b).  An extra coherent gluon line coming out of the polarized nucleon
attaches to one of the dots numbered as (1) and (2).  The propagator with a short bar gives
rise to the SFP at $x_1 = 0$. 
\label{SFP}}
\end{figure}

\begin{figure}[t]
 \begin{center}
  \fig{0.35}{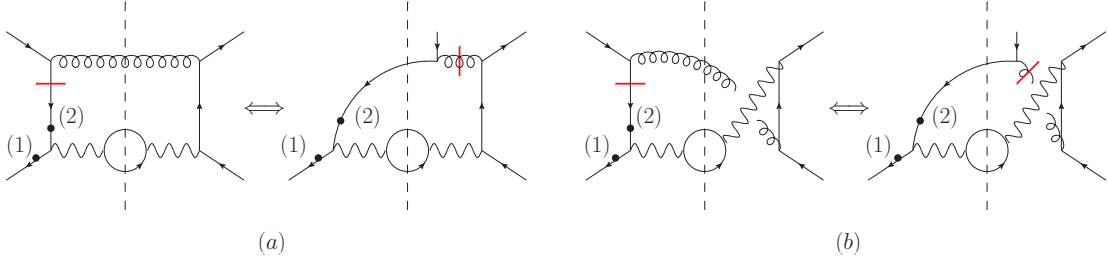}
 \end{center}
 \caption{Cancelling pairs of the diagrams for the SFP contribution.
 An extra coherent gluon line coming out of the polarized nucleon
attaches to one of the dots numbered as (1) and (2).
Two diagrams in the pair (a) and (b) cancel each other.  
\label{cancel}}
\end{figure}

The diagrams giving rise to the SFP are shown 
in Fig. \ref{SFP}, where the quark propagator with a short bar
carrying the momentum $x_1p - x'p'$ (Figs. 4(a) and (b)) and $x_1p + x'p'$ (Figs. 4(c) and (d))
gives the SFP at $x_1=0$.  
Although diagrams in Fig. \ref{cancel} also give rise to the SFP,
they cancels within the pairs (a) and (b).  
This is because in each pair one diagram is obtained from the other by
shifting the position of the cut
and the SFP appears with opposite signs\,\cite{QiuSterman1998,KVY08,KoikeTomita2009}.    
Calculating the diagrams in Fig.~\ref{SFP}, one obtains the SFP contribution as
\bea
\frac{d^4\Delta\sigma^{\rm DY,SFP}}{dQ^2dyd^2\vec{q}_{\perp}} &=&
\frac{\alpha_{em}^2 \alpha_s}{3\pi N S Q^2}
\frac{(-\pi) M_N}{2}\eps^{qpn\Sp} \int\frac{dx'}{x'} \int \frac{dx}{x}
\delta (\shat+\that+\uhat-Q^2) \nn\\
& \times &\sum_a e_a^2 \left[ 
\left\{ \sigmahat^{\rm SFP}_{V, aa} G_{F,a}(0,x) +
\sigmahat^{\rm SFP}_{A, aa} \GFta(0,x) \right\} f_a (x') \right. \nn\\
&& + \left. \left\{ \sigmahat^{\rm SFP}_{V, ag} G_{F,a}(0,x) +
\sigmahat^{\rm SFP}_{A, ag} \GFta(0,x) \right\} G(x')
\right],
\label{SFPcross}
\eea
where the sum for $a$ is over all quark and anti-quark flavors and
the partonic hard cross sections are defined as
\bea
&& \sigmahat^{\rm SFP}_{V, aa} = \frac{2[(\shat-Q^2)^2+\that
Q^2]}{N\shat\that\uhat},\\
&& \sigmahat^{\rm SFP}_{A, aa} =
\frac{2[(\shat-Q^2)^2-\that Q^2]}{N\shat\that\uhat},\\
&& \sigmahat^{\rm SFP}_{V, ag} = \frac{-2[(\shat-Q^2)^2+\that
Q^2]}{(N^2-1)\shat\that\uhat},\\
&& \sigmahat^{\rm SFP}_{A, ag} =
\frac{-2[(\shat-Q^2)^2-\that Q^2]}{(N^2-1)\shat\that\uhat}. 
\eea
Combination of the above results (\ref{HPcross}) and (\ref{SFPcross})
and that for the SGP cross section (see, for example, eq.(29) of \cite{KT071})
gives the complete expression for the leading order (LO)
twist-3 single-spin-dependent cross section 
arising from the quark-gluon correlation functions.

Finally, by making a replacement $\alpha_{em}/(3\pi Q^2) \to \delta(Q^2)$ in 
the Drell-Yan cross section, 
one obtains the cross section formula for the direct photon production (DP). 
Since $x_B$ becomes 0 in this limit, 
all HP's in the DY process transform into SFP's.  
Consequently, we obtain the SFP contribution for the direct photon production as
\bea
E_{\gamma} \frac{d^3 \Delta\sigma^{\rm DP,SFP}}{d^3 \vec{q}} &=&
\frac{\alpha_{em}\alpha_s}{N S} \frac{(-\pi) M_N}{2} \eps^{qpn\Sp} \int
\frac{dx'}{x'} \int \frac{dx}{x} \delta (\shat+\that+\uhat) \nn\\
&\times& \sum_a
\left[ \sum_{b} e_a e_b \sigmahat^{\rm SFP}_{ab} \left\{ \GFa
(0,x) + \GFta (0,x) \right\} f_b (x')\right.  \nn\\
&& \quad + \sum_{b} e_a e_b \sigmahat^{\rm SFP}_{a\bar{b}}\left\{ \GFa
(0,x) + \GFta (0,x) \right\} f_{\bar{b}} (x') \nn\\
&& \left. \qquad +  e_a^2 \sigmahat^{\rm SFP}_{ag}\left\{ \GFa
(0,x) + \GFta (0,x) \right\} G (x') \right] ,
\label{SFPDP}
\eea
where the meaning of $\sum_{a,b}$ is the same as in (\ref{HPcross}) and 
the partonic hard cross sections are given by 
\bea
\sigmahat_{ab}^{\rm SFP} &=& \frac{2(\shat^2+\uhat^2)}{\that^2\uhat} 
+ \frac{2\shat(\uhat-\shat)}{N\that\uhat^2} \delta_{ab},\\
\sigmahat_{a\bar{b}}^{\rm SFP} &=& -\frac{2(\shat^2+\uhat^2)} {\that^2\uhat}
+\left[ \frac{2N\shat}{\uhat^2}+\frac{2(\uhat^2+\shat\that)}{N\shat\that\uhat}\right]\delta_{ab} ,\\
\sigmahat_{ag}^{\rm SFP} &=& \frac{2[N^2\that\uhat-\shat(\shat-\that)]}{(N^2-1)\shat\that\uhat}.  
\eea
In the limit $Q^2\to 0$, the hard cross sections for $G_F$ and $\widetilde{G}_F$
coincide in each scattering channel, 
and thus the SFP functions appear in the combination of $G_F(0,x) + \GFt(0,x)$, 
which was also observed in the
SFP contribution to SSA in the light hadron production $\pup p\to \pi X$\,\cite{KoikeTomita2009}.  
Although we do not understand the reason for this simplification, 
it seems to occur when particles participating in the scattering are all massless.  
This property reduces the number of independent functions contributing to SSA and 
serves greatly for the global analysis of SSA.  
We mention that the sum of the result for the SFP cross section 
in (\ref{SFPDP}) and the result for the SGP cross section (eq.(30) of \cite{KT071})
gives the complete LO twist-3 cross section for the direct-photon production,
$\pup p\to \gamma X$, arising from the quark-gluon correlation functions.

To summarize, we have derived all the HP and SFP contributions to
the twist-3 single-spin-dependent cross section 
for the DY lepton-pair production and the direct-photon production
originating from 
the quark-gluon correlation functions in the transversely polarized nucleon
in the LO QCD.  In the direct photon limit, the HP's are transformed into the SFP's, and
we have observed that 
$G_F$ and $\GFt$ have the common SFP hard cross sections in all channels,
as in the case of the twist-3 cross section for $\pup p\to \pi X$.  
Combined with the existing result for the SGP cross section for these processes, 
this completes the whole twist-3 cross section associated with the quark-gluon correlation functions.  
Forthcoming measurement of the SSA for these two processes at RHIC,
J-PARC, etc, provides us with a valuable opportunity to determine these functions.

\section*{Acknowledgments}
The work of Y. K. is supported in part by the Grant-in-Aid for Scientific Research
No. 23540292 from the Japan Society for the Promotion of Science.


\begin{thebibliography}{10}
\expandafter\ifx\csname url\endcsname\relax
  \def\url#1{\texttt{#1}}\fi
\expandafter\ifx\csname urlprefix\endcsname\relax\def\urlprefix{URL }\fi

\bibitem{review}
U. D'Alesio, F. Murgia, Prog. Part. Nucl. Phys. 61 (2008) 394;\\ 
V.~Barone, F.~Bradamante, A.~Martin, Prog. Part. Nucl. Phys. 65 (2010) 267.  


\bibitem{E7041991}
D.~L. Adams, et~al., Fermilab E704 Collaboration, Phys. Lett. B 261 (1991) 201.

\bibitem{E7041991charge}
D.~L. Adams, et~al., Fermilab E704 Collaboration, Phys. Lett. B 264 (1991) 462.

\bibitem{E7041998}
D.~L. Adams, et~al., Fermilab E704 Collaboration, Nucl. Phys. B 510 (1998) 3.

\bibitem{Star2004}
J.~Adams, et~al., STAR Collaboration, Phys. Rev. Lett. 92 (2004) 171801.

\bibitem{Phenix2005}
S.~S. Adler, et~al., PHENIX Collaboration, Phys. Rev. Lett. 95 (2005) 202001.

\bibitem{Star2008}
B.~I. Abelev, et~al., STAR Collaboration, Phys. Rev. Lett. 101 (2008) 222001.

\bibitem{Brahms2008}
I.~Arsene, et~al., BRAHMS Collaboration, Phys. Rev. Lett. 101 (2008) 042001.

\bibitem{Phenix2010}
A.~Adare, et~al., PHENIX Collaboration, Phys. Rev. D 82 (2010) 112008.

\bibitem{Compass2005}
V.~Y. Alexakhin, et~al., COMPASS Collaboration, Phys. Rev. Lett. 94 (2005)
  202002.

\bibitem{Compass2007}
E.~Ageev, et~al., COMPASS Collaboration, Nucl. Phys. B 765 (2007) 31.

\bibitem{Compass2009}
M.~Alekseev, et~al., COMPASS Collaboration, Phys. Lett. B 673 (2009) 127.

\bibitem{Hermes2001}
A.~Airapetian, et~al., HERMES Collaboration, Phys. Rev. D 64 (2001) 097101.

\bibitem{Hermes2005}
A.~Airapetian, et~al., HERMES Collaboration, Phys. Rev. Lett. 94 (2005) 012002.

\bibitem{Hermes2009}
A.~Airapetian, et~al., HERMES Collaboration, Phys. Rev. Lett. 103 (2009)
  152002.


\bibitem{CS81} J.C. Collins and D.E. Soper, Nucl. Phys. B193 (1981) 381; 
B213 (1983) 545(E). 

\bibitem{CSS85} J.C. Collins, D.E. Soper and G. Sterman,
	Nucl. Phys. B250 (1985) 199. 
	
\bibitem{JMY05} X.~D. Ji, J.~P. Ma and F. Yuan, Phys. Rev. D71 (2005) 034005; 
Phys. Lett. B597 (2004) 299.

\bibitem{Sivers1990}
D.~Sivers, Phys. Rev. D 41 (1990) 83.

\bibitem{Sivers1991}
D.~Sivers, Phys. Rev. D 43 (1991) 261.

\bibitem{Collins2002}
J.~C. Collins, Phys. Lett. B 536 (2002) 43.

\bibitem{Collins93} J.~C. Collins, Nucl. Phys. B396 (1993) 161.

\bibitem{AnselminoEtal2007}
M.~Anselmino, M.~Boglione, U.~D'Alesio, A.~Kotzinian, F.~Murgia, A.~Prokudin,
  C.~T\"urk, Phys. Rev. D 75 (2007) 054032.

\bibitem{AnselminoEtal2009}
M.~Anselmino, M.~Boglione, U.~D'Alesio, A.~Kotzinian, S.~Melis, F.~Murgia,
  A.~Prokudin, C.~T\"{u}rk, Eur. Phys. J. A 39 (2009) 89.

\bibitem{QiuSterman1992}
J.~Qiu, G.~Sterman, Nucl. Phys. B 378 (1992) 52.

\bibitem{QiuSterman1998}
J.~Qiu, G.~Sterman, Phys. Rev. D 59 (1998) 014004.

\bibitem{KanazawaKoike2000}
Y.~Kanazawa, Y.~Koike, Phys. Lett. B 478 (2000) 121.

\bibitem{KanazawaKoike2000E}
Y.~Kanazawa, Y.~Koike, Phys. Lett. B 490 (2000) 99.

\bibitem{JiQiuVogelsangYuan2006DY}
X.~Ji, J.-W. Qiu, W.~Vogelsang, F.~Yuan, Phys. Rev. D 73 (2006) 094017.

\bibitem{Ji:2006br}
  X.~Ji, J.~W.~Qiu, W.~Vogelsang and F.~Yuan,
  Phys.\ Lett.\  B 638 (2006) 178. 


\bibitem{KouvarisQiuVogelsangYuan2006}
C.~Kouvaris, J.-W. Qiu, W.~Vogelsang, F.~Yuan, Phys. Rev. D 74 (2006) 114013.

\bibitem{EguchiKoikeTanaka2006}
H.~Eguchi, Y.~Koike, K.~Tanaka, Nucl. Phys. B 752 (2006) 1.

\bibitem{EguchiKoikeTanaka2007}
H.~Eguchi, Y.~Koike, K.~Tanaka, Nucl. Phys. B 763 (2007) 198.

\bibitem{KT071}
  Y.~Koike and K.~Tanaka,
  Phys.\ Lett.\ B 646 (2007) 232
  [Erratum-ibid.\ {B 668} (2008) 458].

\bibitem{KoikeTanaka2007}
Y.~Koike, K.~Tanaka, Phys. Rev. D 76 (2007) 011502.

\bibitem{KVY08}
  Y.~Koike, W.~Vogelsang and F.~Yuan,
  Phys.\ Lett.\ {B659} (2008) 878. 

\bibitem{KoikeTomita2009}
Y.~Koike, T.~Tomita, Phys. Lett. B 675 (2009) 181.

\bibitem{Koike:2009yb}
  Y.~Koike and K.~Tanaka,
in the proceedings of 17th International Workshop on Deep-Inelastic Scattering and Related Subjects (DIS 2009), Madrid, Spain, 26-30 Apr 2009, 
\verb$http://dx.doi.org/10.3360/dis.2009.209$ 
[arXiv:0907.2797 [hep-ph]]. 

\bibitem{KanazawaKoike2010}
K.~Kanazawa, Y.~Koike, Phys. Rev. D 82 (2010) 034009.

\bibitem{KanazawaKoike2011}
K.~Kanazawa, Y.~Koike, arXiv:1104.0117 [hep-ph].




\bibitem{Beppu:2010qn}
  H.~Beppu, Y.~Koike, K.~Tanaka and S.~Yoshida,
  Phys.\ Rev.\  D {82} (2010) 054005.


\bibitem{Koike:2011ns}
  Y.~Koike, K.~Tanaka and S.~Yoshida,
  arXiv:1104.0798 [hep-ph].


\bibitem{Koike:2011mb}
  Y.~Koike and S.~Yoshida,
  arXiv:1104.3943 [hep-ph].


\bibitem{YZ09} F. Yuan and J. Zhou, Phys. Rev. Lett. {103} (2009) 052001.  

\bibitem{KangYuanZhou2010}
Z.-B. Kang, F.~Yuan, J.~Zhou, Phys. Lett. B 691 (2010) 243.



\end{thebibliography}

\end{document}